\documentclass[journal,a4paper]{IEEEtran}
\usepackage{cite}
\ifCLASSINFOpdf
   \usepackage[pdftex]{graphicx}
   \DeclareGraphicsExtensions{.pdf,.jpeg,.png}
\else
   \usepackage[dvips]{graphicx}
   \DeclareGraphicsExtensions{.eps}
\fi
\usepackage[cmex10]{amsmath}
\usepackage{array}
\usepackage{fixltx2e}
\usepackage{dblfloatfix}
\newcommand{\Sec}[1]{Section~\ref{sec:#1}}
\newcommand{\App}[1]{Appendix~\ref{sec:#1}}
\newcommand{\Fig}[1]{Figure~\ref{fig:#1}}

\newcommand{\Eq}[1]{Eq.\ (\ref{eq:#1})}
\newcommand{\Eqs}[2]{Eqs~(\ref{eq:#1})~and~(\ref{eq:#2})}

\begin{document}
\title{Power Deposition on Tokamak Plasma-Facing Components}

\author{Wayne~Arter,~\IEEEmembership{Member,~IEEE,}
        Valeria~Riccardo~\IEEEmembership{}
        and~Geoff~Fishpool~\IEEEmembership{}
\thanks{W.~Arter, V.~Riccardo and G.~Fishpool are employed by the United
Kingdom Atomic Energy Authority, EURATOM/CCFE Fusion Association,
Culham Science Centre, Abingdon, Oxon. UK OX14 3DB
e-mail: (see http://www.ccfe.ac.uk).}
\thanks{Manuscript received XXXX.}}

\markboth{IEEE Transactions on Plasma Science, submitted}%
{Arter \MakeLowercase{\textit{et al.}}: SMARDDA PFCs}

\maketitle

\begin{abstract}
The SMARDDA software library is used to model plasma interaction with
complex engineered surfaces. A simple flux-tube model of power deposition
necessitates the following of magnetic fieldlines until they intersect geometry taken
from a CAD (Computer Aided Design) database. Application is made to 1) models
of ITER tokamak limiter geometry and 2) MAST-U tokamak divertor designs,
illustrating the accuracy and effectiveness of SMARDDA, even in
the presence of significant nonaxisymmetric ripple field. SMARDDA's
ability to exchange data with CAD databases and its speed of execution
also give it the potential for use directly
in the design of tokamak plasma-facing components.
\end{abstract}


\IEEEpeerreviewmaketitle

\section{Introduction}\label{sec:intro}
\IEEEPARstart{T}{he} problem of economic electrical power generation
using tokamak nuclear fusion continues to generate
new technological challenges, even as the basic issues involved in
magnetically confining plasma become better understood.
Tokamak reactor
designs anticipate long periods, months to years in length,
of continuous plasma discharge
operation at powers of up to $2$\,GW.  
Helium ``ash" produced by fusion that could otherwise prematurely terminate a discharge
must be continually removed from the edge, most likely taking with it a significant
fraction of the total plasma power.
Even in large tokamak experiments, the deposition of 
plasma energy lost from the edge onto likely
construction materials has the potential to do serious structural damage
if restricted to relatively small areas of the size of the plasma
``scrape-off" layer with its $1$\,cm or so thickness.

Plasma-facing component~(PFC) is a generic term for any part of the
tokamak apparatus which could conceivably suffer a significant
amount of power deposition. The main types are limiters, that
are at least in part touching the plasma edge (see \Fig{inrshad3_RZspc}
in \Sec{AFWS} for illustration), and divertors,
into which escaping plasma is channeled to be cooled and/or spread out
(see \Fig{eqsx} in \Sec{MASTU}) and which are expected to be
an essential component of a power-producing reactor.
In either case the physical objects interacting with plasma
consist of panels or tiles made of or at least coated with refractory metal,
protecting complex structures that for example provide active cooling.
Although greater interest attaches to power handling in divertors,
most scenarios for tokamak discharge start-up involve a period
of limiter operation with a comparatively energetic plasma.

This paper shows how a relatively small development of the
software algorithms and modules described for a neutral
beam application in companion Paper~I~\cite{Wa14a}
has allowed the SMARDDA code to examine power deposition in both limiter and
divertor geometries. The limiter studies were in support of ITER~\cite{Ay02ITER},
 whereas divertor designs were tested for the upgrade of the 
MAST spherical tokamak at Culham~\cite{Fi13MAST}. 
Although in
both cases, SMARDDA was used to verify engineering designs produced by others, the speed
of execution of the software would enable it to be more directly involved
the design process.

\subsection{Power deposition problem}\label{sec:problem}
Since the edge plasma is relatively cool, indicative temperature~$10$\,eV,
yet the field is relatively high, indicative level~$5$\,T (in ITER),
the typical ion gyro-radius is very small, $100$\,$\mu$m compared
to a plasma minor radius measured in metres.
Hence, as a consequence
of the individual motions of the charged particles
(ignoring turbulent collective effects), 
the plasma that interacts with PFCs
simply flows along lines of magnetic field.
To spread the power over as large an area as possible, it is
therefore best to arrange the PFCs so that magnetic fieldlines
are at close to tangential incidence on their surfaces, see
formulae for power deposition~$Q$ in \Sec{modelp}.

For a given field alignment and shallow angle of incidence, designing
individual tile surfaces is relatively straightforward~\cite{Mc90Shap}.
However, this is not a complete answer, in that firstly the fieldlines
are are not strictly straight and secondly, the actual installation may
differ substantially from the ideal, notably through the presence of gaps,
and installation tolerances will allow minor misalignments.
The gaps, where power might flow between the tiles/panels, are
necessary to give clearances for installation
and to allow for thermal expansion during discharges. Since the edges of the
tiles, ie.\ those surfaces  adjacent to the designed surfaces, are
at nearly normal incidence to the field, they might have very high levels
of power deposition unless nearby tiles were arranged to shadow them.
However, it is also important to minimise the shadowing of one designed surface
by another as shadowing increases overall average power density.

Further, particularly within a divertor geometry, it may be necessary to
allow for fieldline curvature and ripple, whereas for limiters, it may be
necessary to treat a range of different field alignments, corresponding to different
types of discharges and different times within a discharge. Hence there is the need
for software which can examine detailed designs of sets of tiles, ultimately
defined using a CAD (Computer Aided Design) system, and their
interaction with an accurate 3-D representation of the magnetic field.

\subsection{SMARDDA}\label{sec:SMARDDA}
As described in the companion Paper~I~\cite{Wa14a}, the SMARDDA software
was developed to have in principle all of the features necessary to perform
design-relevant calculations for PFCs. Ancillary software
takes geometry modelled using the CATIA$^{TM}$ design system
and converts it to the ``open" vtk format~\cite{vtkusers} expected
by the main modules, which describes the geometry as triangulated surfaces.
Charged particles, produced by charge exchange
reactions between a neutral beam and plasma in a duct, are tracked in
a magnetic field until they strike the duct walls, and the resulting
power deposition is examined. SMARDDA uses a specially designed
algorithm involving  a special multi-octree type of hierarchical data structure~(HDS)
to speed particle tracking.

However, it is very inefficient to track particle motions directly
when they anyway closely
follow magnetic fieldlines, so development was necessary to solve
the stream- or fieldline equation of motion. In addition, 
new formulae for power deposition are required when it is the local magnetic flux
tube that is responsible for the process, and
new ways are needed to introduce ``particles" into the model to represent
the fieldlines.

Fortunately, the original SMARDDA development
benefitted from the use of Object-Oriented Fortran in its implementation,
which mandates use of  strictly defined and protected objects,
hence a modular structure of code. The concept thus naturally developed
that SMARDDA should become much more a library of object-oriented modules
from which codes for specific purposes could be built as necessary. The
original HDSGEN software written to generate the HDS needed for
SMARDDA ray-tracing exemplifies such a code.

Many codes have been written to track streamlines of fluid flow, and
indeed the facility is available in the freely available
ParaView software~\cite{paraviewguide}
used for the visualisation of vtk files and indeed for much SMARDDA output.
However, the requirement for streamlines to intersect surfaces is more unusual,
and to meet this only literature involving magnetic fields seems
to be relevant. Particle-in-Cell (PIC) codes were discussed in Paper~I,
and as for codes designed specifically to follow fieldlines, most
are either not interested in wall interactions or model them
with idealised geometry. The only documented code with the capability
to treat realistic CAD designs
at the time of the SMARDDA development was Tokaflu~\cite{Mi99Heat},
but this has a number of deficiencies, notably in respect
of computational efficiency. The authors of the general-purpose ISDEP
particle tracking
software~\cite{Ve12ISDE} do not explain how it treats complex geometry.
Very recently a module has been added to the 
magnetic field equilibrium software CREATE
in order to track fieldlines over triangulated geometry~\cite{Ma13Elec}.

The new developments are discussed in the context of the
use of SMARDDA to perform the different modelling tasks, see next \Sec{Introalg}.

\section{General Methods of Calculation}\label{sec:Algo}
\subsection{Introduction}\label{sec:Introalg}
For both limiter and divertor cases, the geometry is logically
divided into two types, the first is the part for which power deposition
is to be calculated (the ``results" geometry) and the second
type is the geometry which protects the edges of the first 
by fieldline shadowing (the ``shadowing geometry"). An
important practical point is where to start the fieldlines, and for
efficiency it seems best to adopt the ``adjoint" concept
from computer graphics, namely to begin the fieldlines on the
results geometry, and test whether they can be followed to the
nominal source of power at the tokamak midplane without
striking the shadowing geometry. Hence the ``results" geometry may
also be referred to as the ``launch" geometry, contrast  Paper~I where
particles are launched to sample analytically defined beamlets
and power is deposited on the shadowing geometry.

\begin{figure}[!t]
\centering
\includegraphics[width=3in]{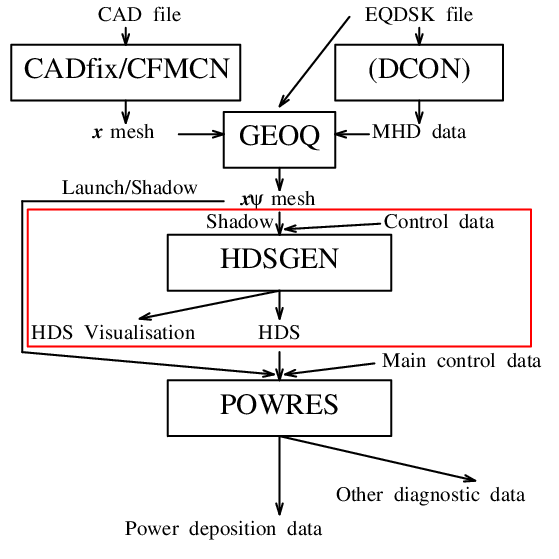}\\
\caption{The flow of data through the SMARDDA
modules for limiter problems.}
\label{fig:flow5t}
\end{figure}

\Fig{flow5t} outlines the flow of data needed to calculate power
deposition on limiter tiles/panels (the data-flow for the divertor
case is very similar). The import of CAD (top left) using
the CADfix$^{TM}$ package
supplied by ITI TranscenData is discussed in detail in Paper~I.
Locally written software converts a CADfix mesh for either
results or shadowing geometry into vtk format 
at the point labelled~``${\bf x}$ mesh" in \Fig{flow5t}, see
\Sec{mesh} below for more details.
``EQDSK" file (top right) denotes a file format which describes a
tokamak magnetic field equilibrium.
Glasser's DCON code~\cite{Gl97Dire} may optionally be
used to check the contents of the EQDSK file, or act as an interface
to other equilibrium field file formats. The magnetic
field~${\bf B}$ is assumed to be axisymmetric,
independent of toroidal angle~$\phi$, and thus can be described
by a (poloidal) flux function~$\psi$, together with a toroidal component
specified by the flux function~$I(\psi)$.

In the limiter case, fieldline calculation takes place
in a coordinate frame aligned with contours of~$\psi$.
Geometry and equilibrium data is combined by the GEOQ code to give
the ``${\bf x}\psi$ mesh" file which has the geometrical information
in flux coordinates, see \Sec{magnetic} below.  The HDSGEN code,
described in Paper~I is shown in
the red box to indicate that it is only required for processing
the shadowing geometry. Although it is apparently thereby implicitly assumed that
the ``results" geometry cannot be struck by a fieldline,
the same tile triangulation can in fact be part of both launch and shadowing geometry.

The fieldline tracing and power deposition calculations are then
performed by the POWRES~code, or the POWCAL code in the case of
divertor calculations. The latter follows fieldlines using cylindrical
polar coordinates in physical space. Features common to both
codes are described in the remainder of this Section, starting with the
power deposition model as this informs subsequent material which
is ordered after the flow of \Fig{flow5t}. Model features and files
specific to the treatment of nonaxisymmetric ``ripple" field are described in \Sec{MASTU}.

\subsection{Model of Power Deposition}\label{sec:modelp}
\begin{figure}[!t]
\centering
\includegraphics[width=1.5in]{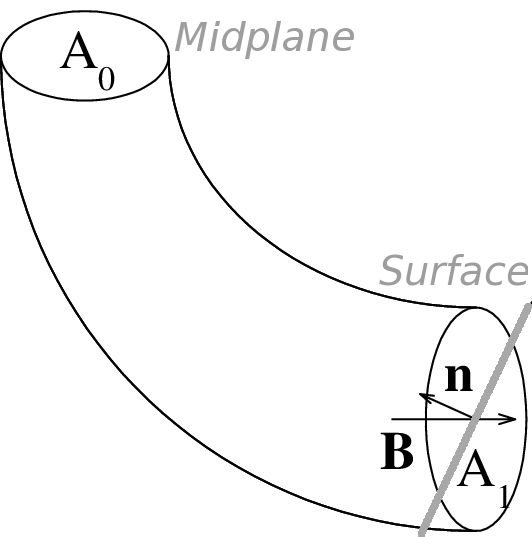}\\
\caption{Flux tube of field~${\bf B}$ connecting the torus midplane
with a surface indicated schematically
in grey with normal~${\bf n}$. The flux tube area is~$A_1$ at bottom
right, at top left it is cut by a horizontal circle with area~$A_0$.}
\label{fig:ftl}
\end{figure}

A simple model of power deposition by plasma in flux tubes is used
in calculations of the tokamak edge. Since it does not seem
to have been fully documented elsewhere, details of
the derivation of the principal formula are presented
in the \App{powerdep}. The basic idea may be explained using
\Fig{ftl}, which shows a flux-tube connecting the tokamak
mid-plane with a physical surface of area~$A_1$.
As it is assumed that particles
follow fieldlines, all power entering the top of the tube strikes
the surface at bottom. Further, the power at the top of the tube
is assumed to fall off exponentially with (major) radial distance at an
empirically determined rate~$\lambda_m$
from the last closed flux surface~(LCFS)~\cite{Go12Heur}. The LCFS
is given by $\psi=\psi_m$
where $\psi_m$ is the value of the poloidal flux where the geometry
touches the plasma, or equal to $\psi$ at the X-point in case of divertor plasmas.
In \App{powerdep}, the power density~$Q$ deposited on
the PFC is shown to vary as
\begin{equation}\label{eq:Qstd}
Q= C_{std} {\bf B}\cdot{\bf n} \exp\left(-\frac{(\psi-\psi_m)}{\lambda_m R_m B_{pm}} \right)
\end{equation}
where~$\psi$ is the flux function value for the tube at the mid-plane,
and the other quantities including the power
normalisation factor~$C_{std}$ (see \App{powerdep})  are fixed for a given equilibrium and geometry.
The formula is found to be a very accurate fit  for suitably chosen~$\lambda_m$
to data from many different tokamak experiments~\cite{Lo07Powe}. 

\subsubsection{Eich formula for Power Deposition}\label{sec:eich}
Eich's formula ~\cite{Ei11Inte} is relevant only to divertor geometries.
It accounts for the spread of power into the ``private flux"
region of the divertor, apparently caused by some kind of collective
plasma behaviour. 
Relative to the previous formula~\Eq{Qstd}, there is
an additional parameter~$\sigma$ to describe the fall-off length
for power deposited in the private flux region, so that
$Q$ varies smoothly across  the surface $\psi=\psi_m$, as
\begin{equation}\label{eq:Q}
Q_E=C_{E} {\bf B}\cdot{\bf n}
\exp \left[ \left( \frac{\sigma }{2\lambda_q } \right)^2 
-\frac{\Delta \psi }{R_m B_{pm} \lambda_q } \right] F_E(\psi)
\end{equation}
where
\begin{equation}\label{eq:FE}
F_E(\psi)=\text{erfc} 
\left( \frac{\sigma }{2\lambda_q } -\frac{\Delta \psi }{R_m B_{pm} \sigma } 
\right) 
\end{equation}
and
\begin{equation}\label{eq:CE}
C_{E}= \frac{F P_{loss}}{4 \pi R_m \lambda_q B_{pm}}
\end{equation}
In the above, $\Delta \psi = \psi-\psi_m$, $\lambda_q$ now denotes
the characteristic decay length outside the private flux region, and
other quantities are as defined in \App{powerdep}. Despite
its formidable appearance~$Q_E(\psi)$ is analytically integrable and 
the normalisation is exact.

\subsection{Meshing and Mesh Refinement}\label{sec:mesh}
\subsubsection{Meshing}\label{sec:submesh}
It is generally sufficient for limiter plasmas to consider
shadowing by adjacent panels only, as confirmed by SMARDDA calculations
with the ITER geometry, see \Sec{AFWS}. This is not necessarily
true for divertor plasmas,
but for the MAST-U work there is the simplification
that the divertor design has twelve-fold symmetry about the major axis.
Hence in both cases, the region to be meshed is reduced to a
small fraction of the total limiter/divertor area, since there are some
$360$~panels in ITER designs (cf.\ over~$100$~tiles in MAST-U divertor).

A major saving both in user and computer time is achieved
by only meshing the surfaces of PFCs. 
For MAST-U,
compared to the typical tile or panel surface dimensions
of~$30$\,cm$\times1$\,m,  the inter-tile gaps are approximately~$2$\,mm.
This implies that to verify absence of power deposition on the tile edges,
triangles with a side of order this length  will be required. A uniform meshing
at~$2$\,mm is excessive in requiring over $10^5$~triangles
per tile, so it is economical to produce a surface
mesh which grades down to this size
only in the critical areas, as elsewhere a $30$\,mm~spacing is sufficient,
at least for exploratory calculations, see \Sec{AFWS} and \Sec{MASTU}.

\subsubsection{Automatic Mesh Refinement}\label{sec:automr}
\begin{figure}[!t]
\centering
\includegraphics[width=1.5in]{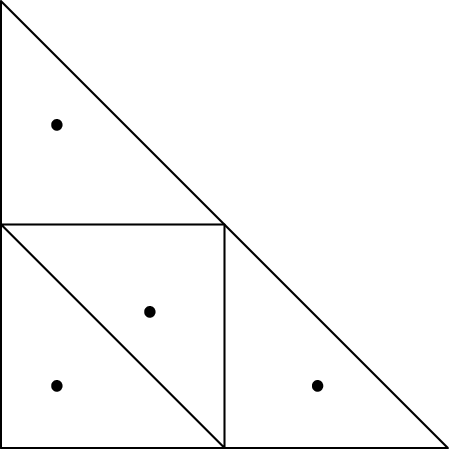}\\
\caption{This illustrates the subdivision of a triangle into
four congruent parts by geometry-conforming points
which is performed automatically by the CFMCN code.}
\label{fig:div4m}
\end{figure}

One unique feature of the CADfix$^{TM}$ package used for the meshing
is its FORTRAN API (Application Programmer Interface). This API has been
used in the development of the CFMCN code capable not only of generating the vtk
files describing triangulations, but also to produce a succession of
refinements automatically for a given triangulation. Each triangle is
separately divided into four as indicated in \Fig{div4m} with the important feature
that the splitting points are geometry-conforming, ie.\ they
lie on the CAD surface and are \emph{not}
just averages of pre-existing points. This ability to produce meshes at
$\times4$ and $\times16$~resolution is most helpful for convergence
studies.
 
\subsubsection{Surface Accuracy}\label{sec:surfn}
For the most part, it is the shadowing of PFC surfaces which is the
main issue. However,
since power deposition $Q \propto {\bf B}\cdot{\bf n}$ with ${\bf B}$
arranged to be nearly perpendicular to~${\bf n}$, if accurate
numerical values are needed, it is necessary to be able to reproduce
the surface normal direction accurately. The economical approach to data
adopted by SMARDDA relies on approximating the tile normal using
the normals of the triangular facetting, rather than say augmenting
the triangulation file with 
${\bf n}$ values extracted from the CAD database. 
The consequent error in $B_n={\bf B}\cdot{\bf n}$ is examined
in detail in \App{geomacc} for a simple configuration of
a toroidal field intersecting a vertical cone of apex angle~$\pi/2-\alpha$.

Assuming that beyond a certain major radius, the cone is
meshed with ``Union Jacks" as explained in \App{geomacc},
then the surface normal computed using plane triangular facets of
toroidal angular extent~$\Delta\phi$
has a component in the toroidal direction equal to~$(\Delta\phi/2) \sin \alpha$.
(The factor of~$\frac{1}{2}$ arises ultimately because two edges define the normal.)
However the logical place to calculate~$B_n$ is at the triangle
barycentre where the geometry-conforming normal is found to contain a 
factor of~$\frac{1}{3}$, leading to an error in~$Q$ scaling as
$\Delta Q \propto(\Delta\phi/6)\sin\alpha$. The ``Union Jack" mesh,
although pleasing to the eye, is actually here very bad for $\Delta Q$.
Even so, the linear scaling of error with mesh-spacing is 
likely to apply for most styles of triangulation, and accounts for
the irregular appearance of $Q$ plots on the (coarse) base meshes
seen in later sections.

\subsection{Magnetic Field Import}\label{sec:magnetic}
For input to GEOQ~(\Fig{flow5t}),
different ITER field distributions are specified using EQDSK files
produced by the CREATE-NL software from the Consortio CREATE~\cite{Al03Plas}, whereas equivalent files
for MAST-U work are produced by the locally written Fiesta code~\cite{gcunn}.
The file format specifies $\psi$ as a set of values
on a regularly spaced grid in~$(R,Z)$. Direct product cubic spline
interpolation of the sample values and their derivatives
using the de Boor package~\cite{deboor}
is used in the obvious way to define the magnetic field at any point
in the gridded region.

Once the flux has been interpolated, it is straightforward to calculate
$\psi_m$ for the limiter plasmas. To determine other parameters
such as~$R_m$ and $\psi_m$ for X-point plasmas,
it is helpful to work in the coordinate
system given by $\psi$ and~$\theta$, poloidal angle measured about
the~O-point in the centre~$(R_{cen},Z_{cen})$ of the plasma. Use of an analytically defined
coordinate such as~$\theta$ reduces these other parameter determinations
to a sequence of 1-D golden-section searches, each in the radial or $\psi$-direction.

Since the ITER calculations work directly with $(\psi,\theta)$ flux
coordinates, it is necessary to calculate the (inverse) mapping functions $R(\psi,\theta)$
and~$Z(\psi,\theta)$. Fortunately for limiter calculations, these are
only needed at such a distance from the X-point that both are well-behaved
functions of~$\psi$. They are calculated point-by-point in much the same
way as the other parameters, ie.\ for each point~$(\psi_i,\theta_j)$ 
of a regular lattice in flux coordinates, a search is conducted along the
radius~$\theta=\theta_j$ to find $(R_i,Z_j)$ such that $\psi(R_i,Z_j)=\psi_i$.
Cubic splines are used throughout to ensure good accuracy, which is
tested  by combining a forward with  a backwards mapping, ie.\ using
$(R,Z)$ evaluated at $(\psi_i,\theta_j)$ as argument to $\psi(R,Z)$,
and verifying that $|\psi-\psi_i|$ is accurate to at least $1$~part in~$10^4$.
\subsection{Fieldline tracing}\label{sec:fline}
In terms of $\psi$ and~$I(\psi)$, the magnetic field
components in cylindrical polars are
\begin{eqnarray}
B_R &=& -\frac{1}{R}\frac{\partial \psi}{\partial Z}  \nonumber \\
B_T &=& I/R \label{eq:bfroma} \\
B_Z &=& \frac{1}{R}\frac{\partial \psi}{\partial R} \nonumber
\end{eqnarray}
where $B_T$ is the toroidal component of field, directed in the $\phi$ coordinate.
The standard fieldline equation is 
\begin{equation} \label{eq:fline} 
\dot{\bf x}=\frac{d {\bf x}}{dt}={\bf B}({\bf x})
\end{equation}
where dot denotes differentiation with respect to pseudo-time~$t$ 
measured along the fieldline. For time independent fields it
may be helpful to think of~$t$ as corresponding to fieldline length.
When flux coordinates are used, \Eq{fline} simplifies to
\begin{equation}
\frac{d \theta}{d \phi}=(1/I) R/J(\psi,\theta)
\end{equation}
where $J(\psi,\theta)$ is the Jacobian of the mapping transformation.

Although the computational costs of solving the
ordinary differential equations~(ODEs) of \Eq{fline} are usually negligible
on modern hardware, it is still important to choose a
numerical algorithm tailored to present requirements, namely
\begin{enumerate}
\item Relatively inexpensive, because $10^3$--$10^5$ or more fieldlines
will need to be computed, corresponding the size of triangulation
of the results geometry.
\item Millimetre accuracy in following fieldlines, corresponding to the
expected accuracy in the position of PFCs subject to thermal expansion
effects.
\item Step sizes such that the fieldline is approximately straight
over one step in~$t$, to ensure accurate geometry intersection.
\end{enumerate}
These requirements are most easily met by a low order scheme with
adaptive timestepping to ensure~(2). Runge-Kutta-Fehlberg~(RKF)
also known as~(\emph{aka}) Embedded Runge-Kutta schemes
with step adaptation by the Shampine-Watts~\cite{Sh79Arto} 
\emph{aka} Cash-Karp algorithm are well documented and relatively easy to implement.
There is however a further restriction on order of accuracy, for
Shampine-Watts relies on a degree of smoothness of the solution~${\bf x}(t)$
at least that of the RKF scheme in order to estimate accurately
the error in the integration. As is well-known, cubic splines
have discontinuous third derivative, hence is sensible to use
third order RKF. Details of the specific schemes implemented
are presented in \App{rkf}.

\subsection{Diagnostics}\label{sec:diagn}
The diagnostics produced by the SMARDDA codes
will be adequately exhibited by plots in \Sec{AFWS} and \Sec{MASTU}.
Output suitable for the open source plotting tool gnuplot is produced
by GEOQ, 
and all plots of 3-D fields unsurprisingly use the vtk format
to be visualised with ParaView~\cite{paraview}.
Moreover, ParaView can perform a wide range of analyses of field data,
which give it
the capability for example, to calculate the total power
deposited on a tile from the contributions of individual elements.

\section{Application to ITER}\label{sec:AFWS}
\subsection{Background}\label{sec:AFWSspecial}
Figure~4 
is produced directly from the ITER
CAD database to illustrate the ultimate starting point.
CAD descriptions of the panels are extracted, and after
defeaturing and repair as explained in Paper~I,
the restricted surface geometry is triangulated as
described in \Sec{mesh}. The resulting meshed geometry
is visualised  in \Fig{panelsinsegc}, where note that each panel
appears as two adjacent geometrical blocks (`semi-panels'), since the central
strip with the wall fixings has been omitted.

The magnetic field geometry  can be understood with
reference to \Fig{inrshad3_RZspc}. Since the tokamak
plasma surface will correspond to a surface of constant
flux to good approximation, limiter contact is made on the
inside of the torus for this equilibrium. The shadowed panel
and the adjacent shadowing panels occupy a volume
such as indicated in \Fig{inrshad3_RZspc}, consequently
only the cross-hatched area need be mapped in flux coordinates.

\begin{figure}[!t]
\centering
\includegraphics[width=3in]{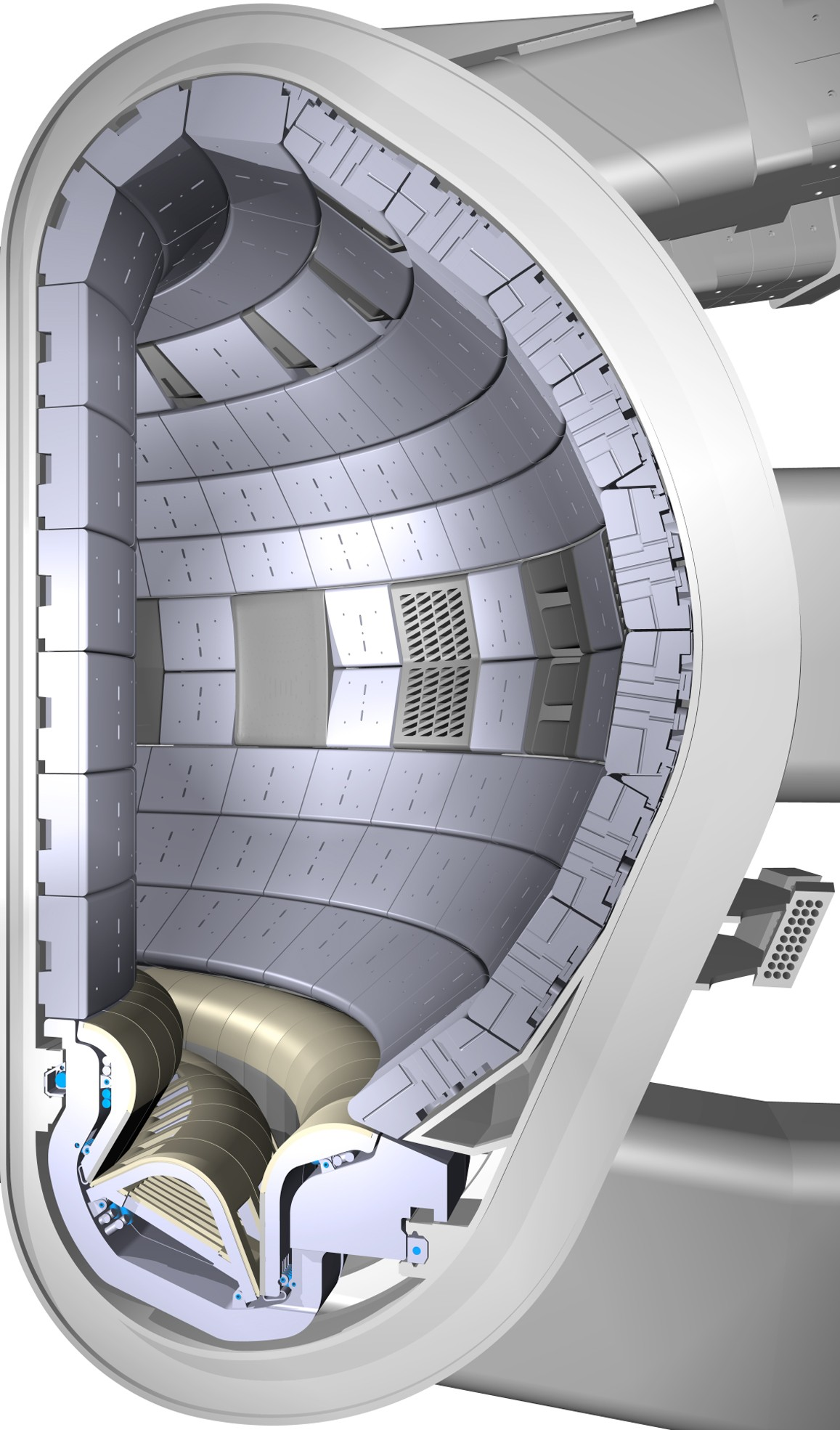}\\
\label{fig:firstwall}
\caption{Vertical cut through detailed ITER model, showing
inside the vacuum vessel, in particular the panels covering the side
and upper walls.}
\end{figure}
\begin{figure}[!t]
\centering
\includegraphics[width=3in]{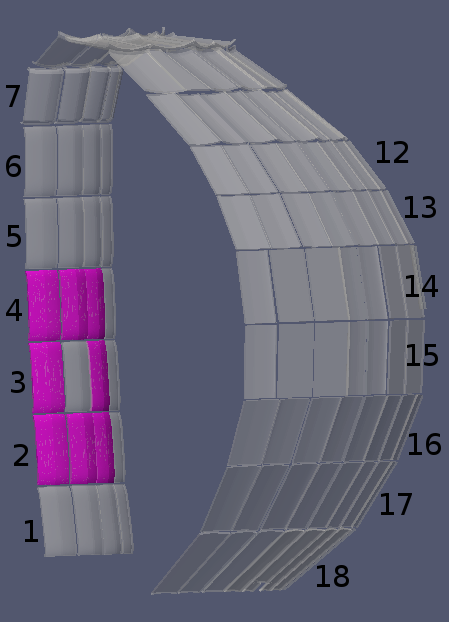}\\
\caption{The surfaces are those of the panels shown in
\protect\Fig{firstwall}, for a $30^o$~segment of the ITER torus, with its ports
blanked off. The numbers refer to horizontal rows of panels.
The panels marked in magenta are tested for their
ability to shadow the panel they surround.}
\label{fig:panelsinsegc}
\end{figure}
\begin{figure}[!t]
\centering
\includegraphics[width=3in]{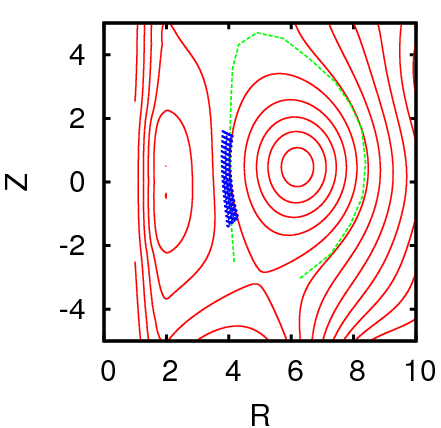}\\
\caption{Contours of magnetic flux~$\psi$ for ITER equilibrium
with $B_T=6.0$\,T, $I_p=7.3$\,MA. The dashed line corresponds to
the inner edge of the first wall in silhouette. The cross-hatching
marks the mapped region containing the point
where the plasma boundary~(LCFS) touches the wall.}
\label{fig:inrshad3_RZspc}
\end{figure}

\subsection{Illustrative Results}\label{sec:AFWSresults}
\begin{figure}[!t]
\centering
\includegraphics[width=3.25in]{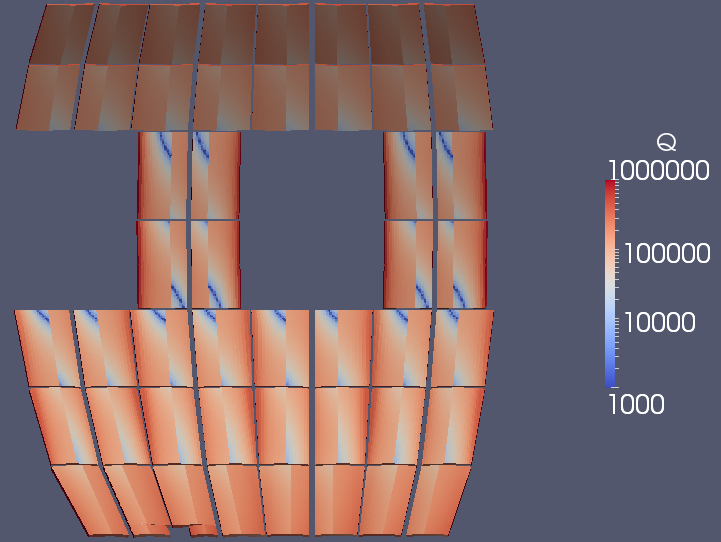}\\
\caption{The outer panels, rows $12$ to~$18$ of the ITER model are shown
``painted" with power~$Q$, ignoring shadowing effects.
Equilibrium with $I_p=7.5$\,MA and $B_T=6.0$\,T,
$P_{loss}=5$\,MW and  $\lambda_m=90$\,mm.}
\label{fig:x2-2lc}
\end{figure}

The results presented are chosen to give a flavour of the
effort needed to thoroughly verify and validate SMARDDA for
limiter work, as well as demonstrate potentially useful capabilities
of the codes. One such is the ability, having calculated
fields~${\bf B}$ and~$\psi$ for each surface, to colour each triangle
with the value of~$Q$ given by \Eq{Qstd}. This enables  the
SMARDDA calculation of~$Q$ to be tested against direct evaluation
of the formula~\Eq{Qstd} using the ParaView calculator,
which has its own method for finding~${\bf n}$. The resulting
$Q$~distribution is calculated at negligible computational
cost since no shadowing is performed, but can give insight
into surfaces most at risk of overheating, see \Fig{x2-2lc}.

\begin{figure}[!t]
\centering
\includegraphics[width=2.5in]{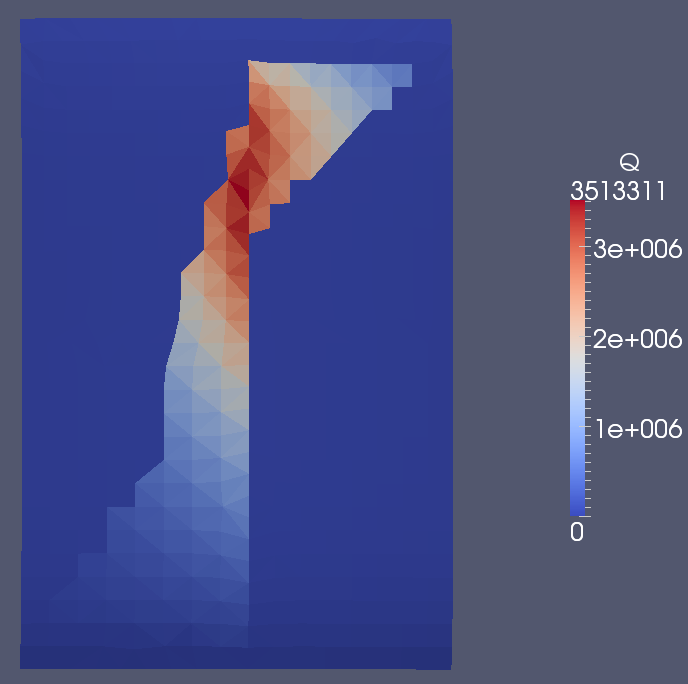}\\
\caption{Power deposited on the central (semi-)panel accounting
for shadowing by its $8$~nearest neighbours. The
base meshing of both ``results" and shadowing geometry
has been used.}
\label{fig:ei1by1c}
\end{figure}

Since the ability to treat CAD is key, much interest attaches
to the influence of the discretisation of the geometry on the
power deposition results. \Fig{ei1by1c} is indicative of the
results produced on the base mesh, ie.\ the mesh produced 
directly using the CADfix mesher. For the ITER panels, the
nominal mesh length is~$30$\,mm, giving $3350$~surface triangles
on the launch geometry.
This translates directly into number of fieldlines followed,
of which $382$~escape past the shadowing panels and produce
the power distribution shown in \Fig{ei1by1c}, for which
$P_{loss}=7.5$\,MW and  $\lambda_m=50$\,mm.

\begin{figure}[!t]
\centering
\includegraphics[width=2.5in]{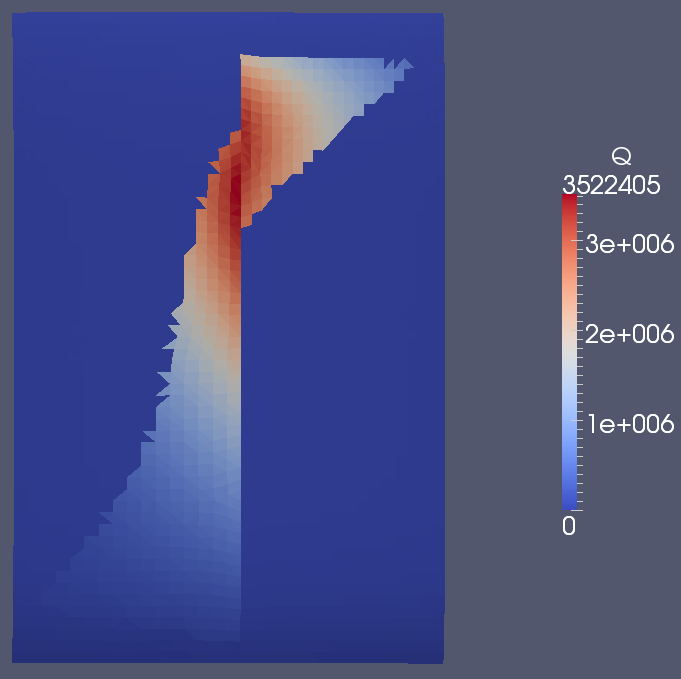}\\
\includegraphics[width=2.5in]{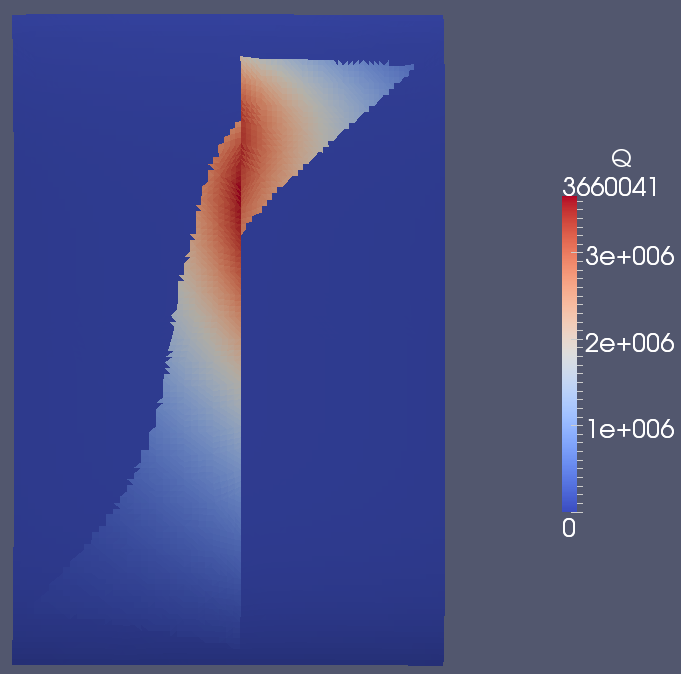}\\
\caption{Power deposited on the central semi-panel accounting
for shadowing by its $8$~nearest neighbours. The
base meshing of the ``results" geometry is successively
refined~$\times4$ (top) and~$\times16$ (bottom).}
\label{fig:ei23by2}
\end{figure}

In addition to successful comparisons with streamlines
produced by ParaView,
there was further detailed examination of the fieldline
integration algorithm, to understand how control of the local
error limits the global error. Thus it emerged that the fieldlines
are so close to being straight in flux coordinates
that as few as $10$~steps might be needed to get past the
shadowing tiles. In any event since here the relevant computed
properties are the area of shadowing and the $Q$~dependence,
the demonstration that changing the integration tolerance~$\epsilon_r$
makes no appreciable to these properties, suffices to 
prove acceptable error control. Indeed, \Fig{ei23by2} is
unchanged if $\epsilon_r$ is increased from $10^{-6}$ to~$10^{-4}$.
\Fig{ei23by2} shows deposition results for a shadowing geometry
refined up to~$\times 16$ relative to the base meshing drawn in \Fig{ei1by1c}.
As the number
of launch points is increased,  evidently peak power deposition
and $Q$ distribution change little.

The total computation time for most refined calculation was~$7$\,s
on an AMD Athlon 64~X2 dual core processor.
The tracking calculation took $3.73$\,s, during which time
$53\,600$ fieldlines were tested for intersection 
with the $289\,040$ triangles in shadowing geometry
(and  $6\,050$ escaped). The approximate time for each fieldline
calculation was therefore~$70\,\mu$s. With approximately~$10$
steps per fieldline, this gives a cost of $7\,\mu$s per straight
track, a figure which compares very favourably with similar
numbers found Paper~I for the duct problem which had only $2\,146$ triangles.
It can be concluded that use of the HDS can make the cost
of geometry intersection tests almost independent of geometry
complexity.

\begin{figure}[!t]
\centering
\includegraphics[width=3.25in]{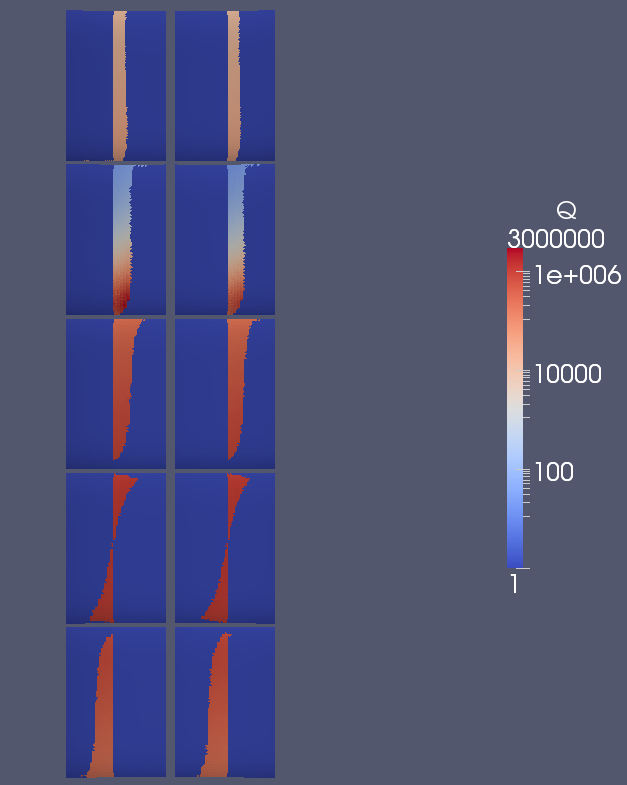}\\
\caption{Power deposited on the panels in rows $2$ to~$6$
assuming that each is shadowed by its $8$~nearest neighbours.}
\label{fig:eq5r2-6}
\end{figure}

Lastly \Fig{eq5r2-6} is an example of a shadowing study used to
check the normalisation of~$Q$, but also of direct
relevance to the designer, for an ITER start-up phase (quasi-)equilibrium
at $t=7.34$\,s with $I_p=3.11$\,MA and wall plasma safety factor~$q_{wall}=8.73$.
$P_{loss}=3.17$\,MW and  $\lambda_m=146$\,mm.
The line where panel illumination changes
from right-handed to left-handed is where the plasma touches the geometry.
Computations were also conducted with $14$~panels that demonstrated
that shadowing by second nearest neighbour panels was generally unimportant,
except in the case of the ports in rows~$14$ and~$15$.
These results led onto studies of the effect of small
panel misalignments on power deposition, using yet another SMARDDA code VTKTFM
to manipulate the vtk files directly to displace and rotate ``results"
and ``shadowing" geometries.

\section{Application to MAST-U}\label{sec:MASTU}
\subsection{Special features for MAST-U}\label{sec:MASTUspecial}
\begin{figure}[!t]
\centering
\includegraphics[width=3.25in]{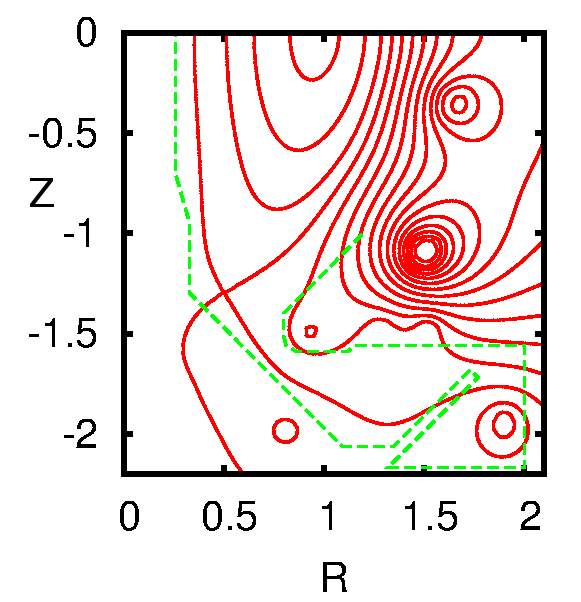}\\
\caption{Contours of magnetic flux~$\psi$ for lower half of MAST-U
Super-X equilibrium
with $B_T=0.64$\,T, $I_p=1.0$\,MA. The dashed line corresponds to
the inner edge of the first wall in silhouette.}
\label{fig:eqsx}
\end{figure}

\begin{figure}[!t]
\centering
\includegraphics[width=3.25in]{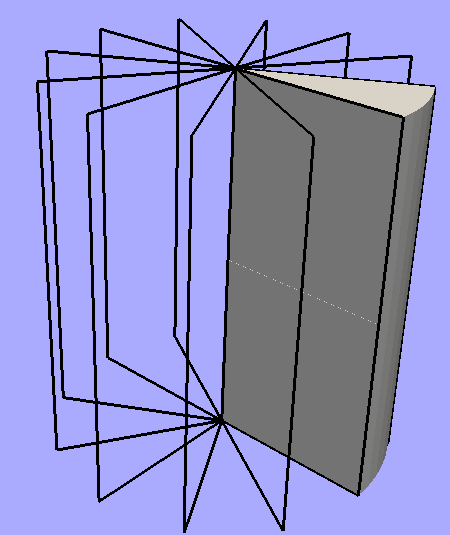}\\
\caption{Sketch of MAST toroidal field (TF) coils, shown as black lines,
with current feeds omitted for clarity.
Only the geometry in the lower half of a $30^o$ segment (marked)
need be modelled.}
\label{fig:mastsketch}
\end{figure}

The application to divertors is more challenging because the
X-point topology means there is no simple 2-D mapping from space to flux-based
coordinates, see eg.\ \Fig{eqsx} near $R=0.5$, $Z=-1.3$.
Moreover the fact that the external toroidal confinement
field is produced by a set of discrete coils manifests itself
as a ripple with period in the toroidal direction proportional
to $1/N_s$, where the number of coils $N_s=12$ for MAST, see
\Fig{mastsketch}. Tokamak toroidal field (TF) coils are generally designed so
that this ripple is negligible in the central plasma region, but
the ripple requires special treatment in MAST-U,
because existing physical constraints virtually force the
divertor into a space near the TF conductors, see \Sec{ripple}.
However, since twelve-fold symmetry extends to the divertor geometry
it is sufficient to work with a $30^o$ segment, see \Sec{symmetry}.

\subsubsection{Ripple field}\label{sec:ripple}
The treatment of the ripple requires provision of all three components of 
the magnetic field generated by a set of current loops such as sketched
in \Fig{mastsketch}. Indeed preliminary power deposition calculations
were performed using the magnetic field from exactly this current configuration.
However detailed investigations which will be reported elsewhere, revealed
that it was important to account both for the finite width of the conductors
and for the current feeds.
In either case, the magnetic field is provided to SMARDDA as sample values
on a uniform grid in $(R,\phi,Z)$~coordinates covering the volume
sketched in \Fig{mastsketch}. (The distinction between the use of~$\phi$
and $\zeta$ for angular coordinate in the toroidal direction is explained
in \App{geomacc}.)

As in \Sec{AFWS}, ${\bf B}$ component values at an
arbitrary point~$(R,Z)$ are calculated by direct product
cubic spline interpolation between the supplied mesh values.
However, for interpolation in toroidal angle~$\zeta$, the
periodicity makes optimal the use of a Fourier series
representation. The actual coil geometry lacks reflectional symmetry
so the vacuum magnetic field dependence on~$\zeta$ has to be
written
\begin{eqnarray} \label{eq:f2dim}
B_{i v} = B_{i0}&+&\Sigma_{m=1}^{m=N_m} B_{ism} \sin m\xi \nonumber \\
&+& \Sigma_{m=1}^{m=N_m} B_{icm} \cos m\xi, \\ 
i&=& R,\;\;Z,\;\;\zeta \nonumber
\end{eqnarray}
where $N_m$ is determined by the data sampling rate,
and the scaled angle $\xi=N_s \zeta$.
Since $32$~samples in $\xi$ are provided, $N_m=16$ provides
an exact representation of the data at uniformly spaced
intervals in $\xi$ or~$\zeta$.

\begin{figure}[!t]
\centering
\includegraphics[width=3.25in]{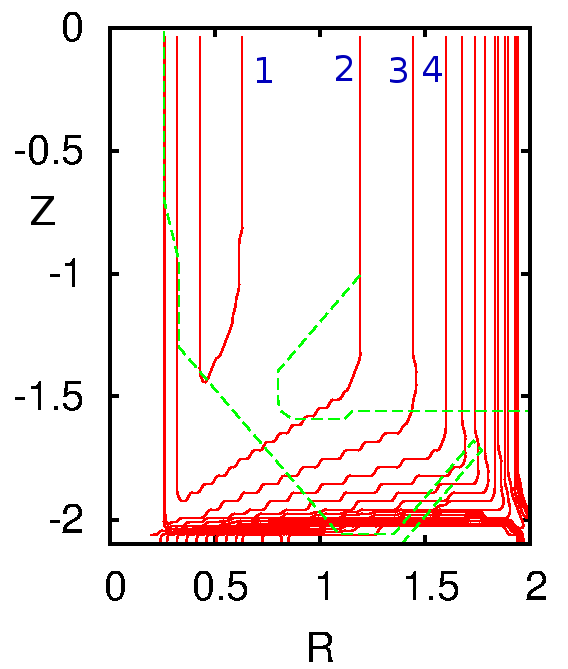}\\
\caption{Contours of~$N_{m\zeta}$ needed
to reproduce field component~$B_{\zeta v}$ to a given accuracy~$\epsilon_m=10^{-6}$
as a function of~$(R,Z)$. The four lowest contours are labelled with
their $N_{m\zeta}$~values.}
\label{fig:cpt3l}
\end{figure}

The Fourier expansion coefficients $B_{icm}$ and~$B_{ism}$ in \Eq{f2dim} are straightforwardly
evaluated using fast Fourier transforms, and the mode spatial
dependences examined. At each~$(R,Z)$ the minimum number of angular modes~$N_{mi}$
necessary to reproduce the samples of field component~$B_{i v}$
to within a specified relative tolerance~$\epsilon_m$ may be computed, and contour plots
such as \Fig{cpt3l} produced.
The field design ensures that the hottest plasma occupies
a region where the field is close to axisymmetric, so that $N_{m\zeta}\geq 2$
is required only for the extremities of the divertor region.
The local axisymmetry also makes it easy to normalise ${\bf B}_v$ consistent with~$I$
used in the magnetic equilibrium calculation.

Accounting for ripple, the fieldline integration \Eq{fline} becomes,
since the fieldlines are unchanged when each component is multiplied by
an identical function of position, 
\begin{eqnarray}
\dot{R} &=& -\frac{\partial \psi}{\partial Z}+ RB_{Rv} \nonumber \\
\dot{Z} &=& \frac{\partial \psi}{\partial R} +R B_{Zv}\label{eq:f3dr} \\
\dot{\xi} &=& N_s B_{\zeta v} \nonumber
\end{eqnarray}
and is solved in the natural coordinates~$(R,Z,\xi)$.
To solve \Eq{f3dr}, it was found
computationally efficient to selectively mask out the higher mode numbers
of the vacuum field~${\bf B}_v$ depending on
position. Using a relative tolerance of~$\epsilon_m=10^{-6}$
to determine the mask, the cost of 
following fieldlines in the divertor region could be reduced by a factor of two.
Further economy was achieved by tracing fieldlines only as far as the plane
$Z=-1.29$\,m below the centre of MAST, where it is certain that they are able
to connect power from the midplane.

As in \Sec{AFWSresults}, both analysis and single fieldline numerical
calculation were
used to establish conservative parameter values for
fieldline tracing of sufficient, millimetre accuracy. In particular
it was found preferable to use an absolute tolerance~$\epsilon_a$ for
fieldline integration. This tolerance was
then used to calculate power deposition profiles, which were found to
be invariant under order of magnitude increases in~$\epsilon_a$ for example.
\subsubsection{Symmetry}\label{sec:symmetry}
\begin{figure}[!t]
\centering
\includegraphics[width=3in]{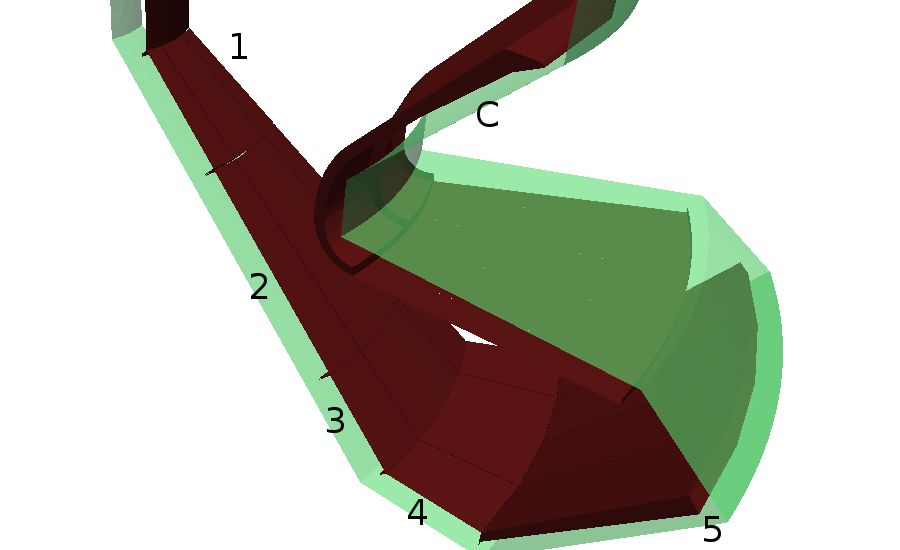}\\
\caption{The dark surfaces are those of the geometry modelled 
for a $30^o$~segment of a MAST-U divertor design.
An enveloping ``bean-can" to catch any leakage of fieldlines through the tiles
is shown as a lighter halo.
The numbers refer to horizontal rows of tiles in the divertor and
``C" labels the coil armour shadowing geometry.}
\label{fig:mastorilc}
\end{figure}

\Fig{mastorilc} shows a representative one-twelfth
of the MAST-U geometry in and above the divertor, indicating the
locations of tiles T1--T5.  Testing for fieldline intersections
with the complete~$360^o$ device  is achievable using this $30^o$~sector, by
applying the periodic condition ${\bf x}(R,Z,\xi+2\pi)={\bf x}(R,Z,\xi)$
when fieldlines exit the sector.
\subsection{Illustrative Results}\label{sec:MASTUresults}
The MAST-U calculations represent the most demanding application of SMARDDA
to date since the fieldlines may be long in terms of both number of circuits
around the vertical axis and the number of steps often exceeds~$500$
(average~$100$). Further, although the
geometry is physically smaller, it contains as much detail as in \Sec{AFWS}.
An important design goal of MAST-U is the ability to handle the
``Super-X" field configuration~\cite{Fi13MAST} as well as the usual X-point arrangement,
which in practice means the ability to
handle intermediate configurations wherein power can be deposited
on any of tiles T1 to~T5.
All equilibria have $B_T=0.64$\,T and $I_p=1.0$\,MA, with
additional shaping by varying the current in external 
poloidal field coils, the location of which
can be deduced from the extrema of~$\psi$ in say, \Fig{eqsx}.

Detailed, realistic predictions of power deposition 
are of interest principally to MAST-U designers. With
ease of exposition in mind, scrape-off layer width is
taken as $\lambda_m=10$\,mm, rather than the more realistic~$3$\,mm 
in order to make the plots of power deposition easier to visualise.
The total power loss is set to be~$P_{loss}=1$\,MW,
a quantity chosen on the grounds that it is easily scalable
to other values, but it is not actually  
a physically expected value.
In this context of easier exposition, deployment of SMARDDA to examine the uniformity
of power deposition will be described in \Sec{MASTUp}, and use
of the software to identify issues
raised by preliminary examination is documented in
\Sec{MASTUident}, immediately below.

\subsubsection{Identification of Issues}\label{sec:MASTUident}
\begin{figure}[!t]
\centering
\includegraphics[width=3.25in]{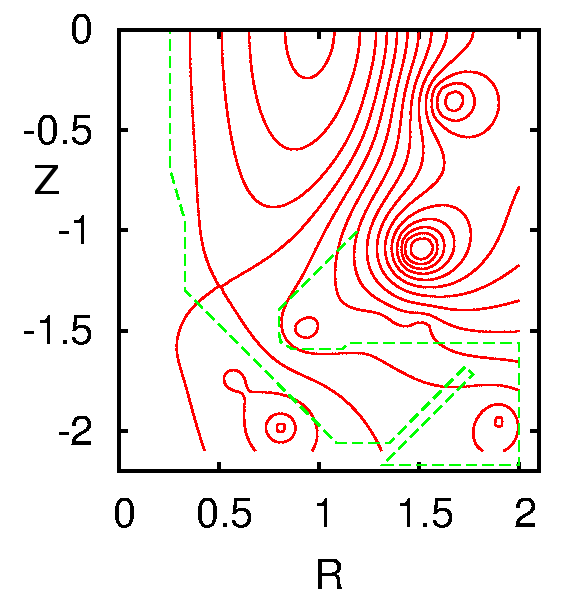}\\
\caption{Contours of magnetic flux~$\psi$ for MAST-U
intermediate equilibrium.}
\label{fig:eq4}
\end{figure}

\begin{figure}[!t]
\centering
\includegraphics[width=2.0in]{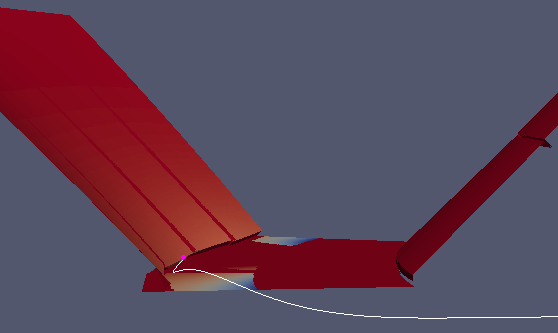}
\includegraphics[width=1.25in]{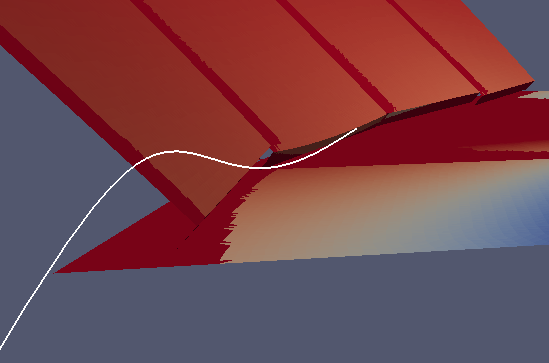}\\
\caption{MAST-U intermediate equilibrium.
The plots of views from two slightly different directions show
how a fieldline (in white)  may get into the T4/T5 gap in
one design.}
\label{fig:odear}
\end{figure}

The equilibrium of \Fig{eq4} shows that the plasma scrape-off layer,
approximated by the contour passing through the X-point, is
located close to the boundary between tiles T4 and~T5. Plots of~$Q$
deposition showed that there was power
deposited on the lower edge of T5, at potentially significant levels
because of near-normal fieldline incidence. This would be unexpected in an
axisymmetric field, but \Fig{odear} shows that the field ripple
is sufficiently large at the T4/T5 gap to allow it. Further
examination verified that the total power deposited on the edge of
one T5 tile was less $0.1$\,\% of the total loss, ie.\ negligible.

Calculations of power deposited by a standard X-point plasma
on tiles T2 and T3 showed a similar failure of shadowing. This
issue was resolved by making minor modifications to the design.

\subsubsection{Distribution of Power Deposition}\label{sec:MASTUp}
\begin{figure}[!t]
\centering
\includegraphics[width=3in]{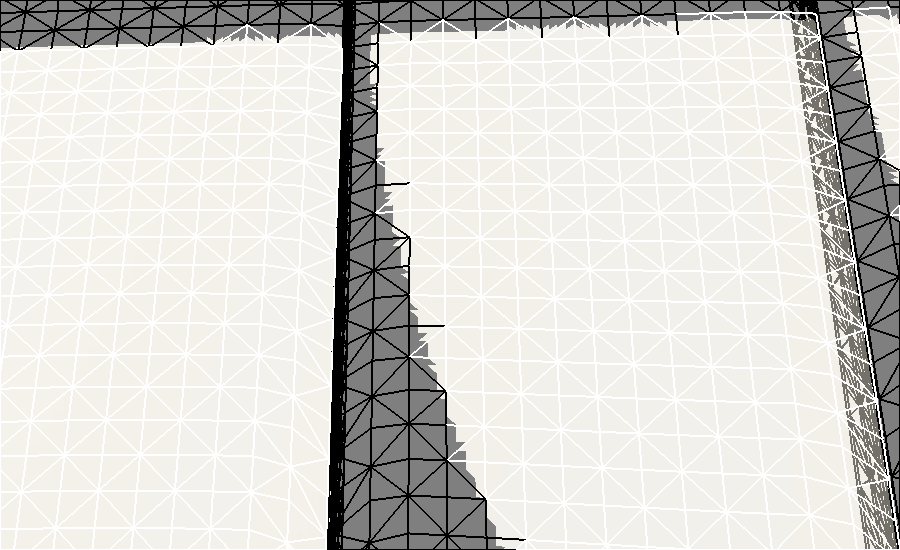}\\
\caption{Power deposition shadows for Super-X equilibrium. Expanded
view of upper region of T5 surfaces showing a
superposition of illuminated areas from calculations
with a coarse mesh and the same mesh with $16\times$~refinement.
The coarse mesh ilumination is marked by the edges of the mesh
triangles (black for shadow) whereas the finer mesh illumination
is indicated by solid white.}
\label{fig:uushad}
\end{figure}

Observe from \Fig{uushad} that overlaying the $Q$ distributions
for different meshings of the
geometry shows changes only in $Q$ at the cutoff boundaries
which clearly correspond to the triangle subdivision algorithm.
Once this had been verified and the fieldline tests mentioned in \Sec{ripple} had been passed,
detailed calculations were performed to demonstrate uniformity of power
deposition on tiles as the equilibrium properties were varied.

\begin{figure}[!t]
\centering
\includegraphics[width=3in]{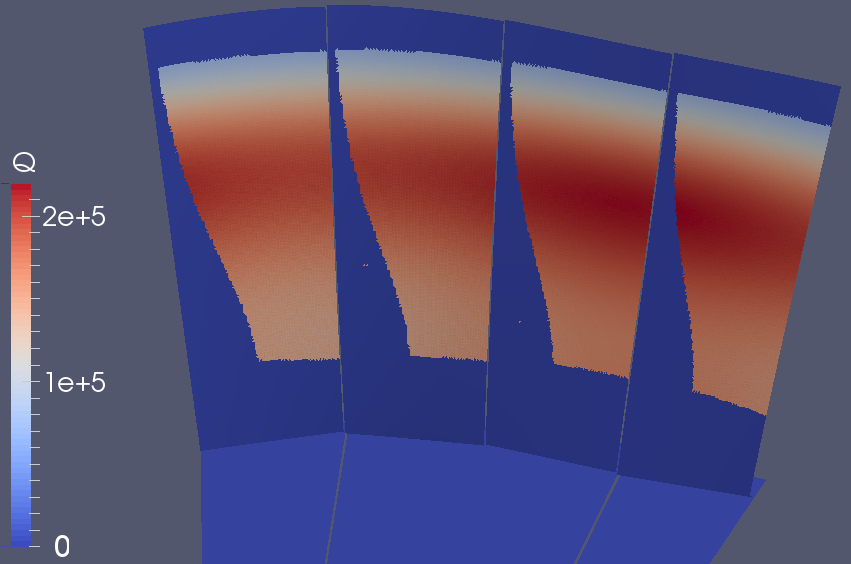}\\
\caption{Distribution of power deposited on T5 tiles for MAST-U Super-X equilibrium,
with revised tile design, $\lambda_m=10$\,mm.}
\label{fig:l0plus}
\end{figure}

\Fig{l0plus} is indicative of the results obtained. The variation in total
power deposited varies
from tile to tile by under~$5$\,\%.
The total computation time for this calculation was just over
an hour, viz.\ $4\,000$\,s
spent testing $67\,776$ fieldlines for intersection 
with the $29\,063$ triangles in the shadowing geometry
from which  about half of the fieldlines escape.
The approximate time for each fieldline
calculation was therefore~$60$\,ms. The three orders of magnitude increase
in cost relative to the ITER calculations is accounted for
by much longer fieldlines' forming a higher percentage of the
total, and the large increase in the number of calculations
needed to evaluate~${\bf B}$.


\begin{figure}[!t]
\centering
\includegraphics[width=3in]{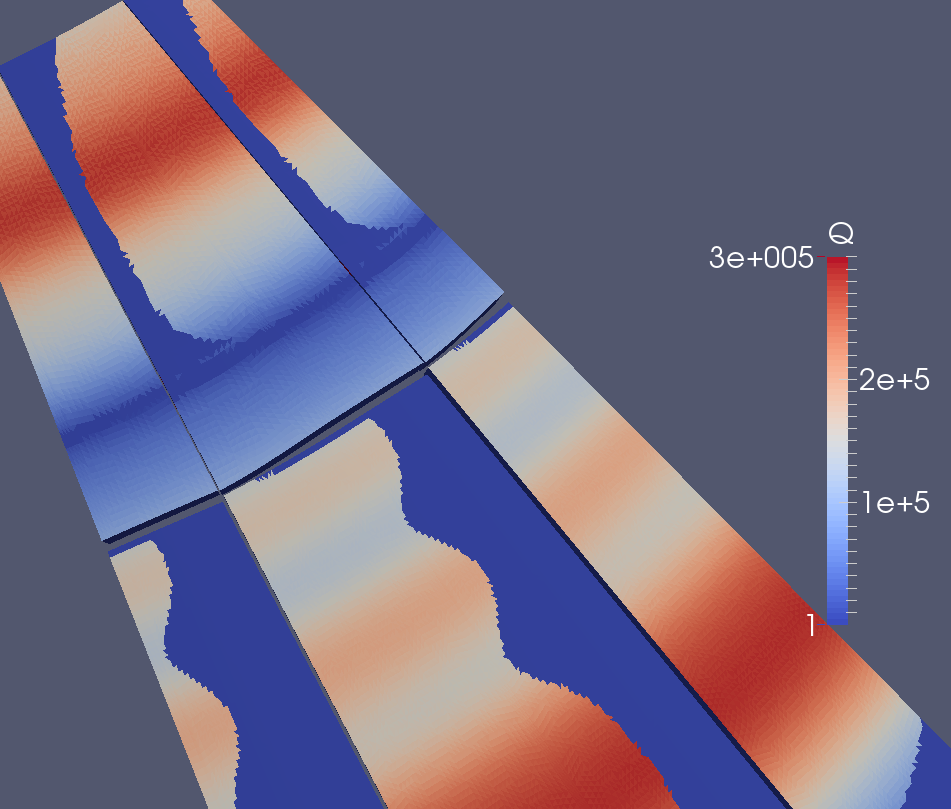}\\
\caption{Distribution of power deposited on tiles T1~and~T2
for MAST-U X-point equilibrium
using Eich formula with overlarge $\sigma=\lambda_q=10$\,mm.
The immediate neighbourhood of the horizontal band without
power on (the upper tile) T1 is an artefact.}
\label{fig:res12gra}
\end{figure}

Lastly, \Fig{res12gra} shows how the Eich formula gives finite power
deposition within the private flux region, for tiles T1 and T2.
As in the case of the simpler formula, the power fall-off lengths have
been deliberately exaggerated for ease of exposition,  which 
here has the useful side-effect of maximising the area
of illumination, thereby exposing any areas where power
is deposited anomalously.

\section*{Acknowledgment}
Valuable input from and discussions with R.~Mitteau of ITER,
G.~Saibene of F4E and P.~J.~Lomas of CCFE  are gratefully acknowledged.
Magnetic equilibrium fields for ITER
and MAST-U respectively were calculated and
supplied by M.~Mattei of Consorzio CREATE, Second University of Naples,
and G.~Cunningham of CCFE. D.~Taylor of CCFE helped
with the MAST ripple field.
CAD files for ITER and MAST-U geometry respectively were supplied by G.~Saibene
of F4E and N.~Richardson of CCFE.\\
ITER related work was funded by F4E contract F4E-OPE-148.
This work was funded by the RCUK Energy Programme grant number EP/I501045
and the European Communities under the contract of Association between
EURATOM and CCFE.
To obtain further information on the data and models underlying this
paper please contact PublicationsManager@ccfe.ac.uk.
The views and opinions expressed herein do not necessarily reflect
those of the European Commission.

\appendices
\section{Physics Model}\label{sec:powerdep}
\Fig{ftl} shows part of a flux tube,
which has area~$A_1$ at the bottom right end and at top left is cut by
horizontal circle of area~$A_0$ at the midplane. The poloidal magnetic field
has strength~$B_1$ at~$A_1$ and is vertical corresponding
to a poloidal component~$B_0$ at~$A_0$.
Since flux is
conserved, $\nabla \cdot {\bf B}=0$, it follows that
\begin{equation}
B_0 A_0 = B_1 A_1
\end{equation}

Now, suppose that at the midplane, particles with
energy $E_s=m v_s^2/2$ enter the tube, say $N_p$ in a unit time interval.
In steady state, they leave the tube at bottom at the same rate
by particle conservation.
Their equation of motion is
\begin{equation}
m \frac{d {\bf v_s}}{dt}=-e ({\bf \nabla\Phi} + {\bf v_s} \times {\bf B})
\end{equation}
where ${\bf -\nabla\Phi}$ is the electric field, $m$ is particle mass and $e$
is the particle charge. Energy conservation for each particle gives
\begin{equation}
\frac{m v_s^2}{2}+ e\Phi = \mbox{const.}
\end{equation}
Hence in the absence of an electric field, particle energy is conserved.
Indeed, provided that equal charges of ions and electrons pass down the
tube, energy is conserved even for $\nabla\Phi \neq {\bf 0}$. Note that
the model neglects collective particle (`fluid') effects and does not
for example, allow for loss of particles through the tube walls via finite 
gyro-radius effects.
It follows that
the particle energy leaving the tube in unit time at~$A_1$ is
$E_t=\Sigma_{s=1}^{N_p} E_s$, and the power density at~$A_1$ is therefore
\begin{equation}
\frac{E_t}{A_1}=\frac{E_t B_1}{A_0 B_0}
\end{equation}
However, this power is spread out when the tube strikes a surface
(shown dashed in \Fig{ftl}) inclined to the flux-tube, so the
density is reduced by a factor $|{\bf B}\cdot{\bf n}|/B_1$, hence
the power density on the surface is
\begin{equation}
Q=\frac{E_t}{A_0} \times \frac{{\bf B}\cdot{\bf n}}{B_0}
\end{equation}
written to separate out a factor representing power density at midplane.

Suppose that the total power lost from the plasma is lost at the
midplane and is $P_{loss}$, and further that there is an exponential
fall-off in the lost power density with distance from the
plasma boundary at a rate~$\propto \exp(-\Delta R/\lambda_m)$,
where $\Delta R=R-R_m$.
If the maximum power density is~$Q_0$, then 
\begin{equation}
\frac{E_t}{A_0} = Q_0\exp(-\Delta R/\lambda_m)
\end{equation}

To relate $Q_0$ to $P_{loss}$, imagine that
the outer midplane boundary of the plasma lies at a major radius
of~$R_m$.  Integrating over the exterior of the circle~$R=R_m$
gives
\begin{equation}
\int_0^{\infty} d(\Delta R) \int_0^{2\pi} R_m d\theta Q_0 \exp(-\Delta R/\lambda_m) =P_{loss}
\end{equation}
or
\begin{equation}\label{eq:ploss}
Q_0 = \frac{P_{loss}}{2 \pi R_m \lambda_m}
\end{equation}

It is convenient to use coordinates based on the flux~$\psi$,
making the Taylor series approximation
\begin{equation}
\Delta R \approx \frac{\partial \psi}{\partial R} \Delta \psi
\end{equation}
where $\Delta \psi=\psi-\psi_m$.
At the midplane, the poloidal component of ${\bf B}$ is given
by
\begin{equation}
B_{pm} = \frac{1}{R}\frac{\partial \psi}{\partial R}
\end{equation}
hence
\begin{equation}
\frac{\Delta R}{\lambda_m} \approx \frac{\Delta \psi}{\lambda_m R_m B_{pm}}
\end{equation}
Combining the above formulae, noting that $B_0=B_{pm}$,
gives the power density deposited on
the surface with normal~${\bf n}$ as 
\begin{equation}
Q= \frac{F P_{loss}}{2 \pi R_m \lambda_m B_{pm}} {\bf B}\cdot{\bf n} \exp\left(-\frac{(\psi-\psi_m)}{\lambda_m R_m B_{pm}} \right)
\end{equation}
where $F$ is the fraction of lost power going down the flux-tube,
normally $F=\frac{1}{2}$ to allow for an equal amount of
power's being lost in the opposite direction.
Hence $Q$ is determined as a function of flux~$\psi$
by two parameters which are in principle freely
specifiable by the modeller, viz.\ $P_{loss}$ and~$\lambda_m$, and quantities
evaluated by the GEOQ code, namely $R_m$, $B_{pm}$ and~$\psi_m$.

\section{Geometrical Accuracy}\label{sec:geomacc}
\begin{figure}[!t]
\centering
\includegraphics[width=2in]{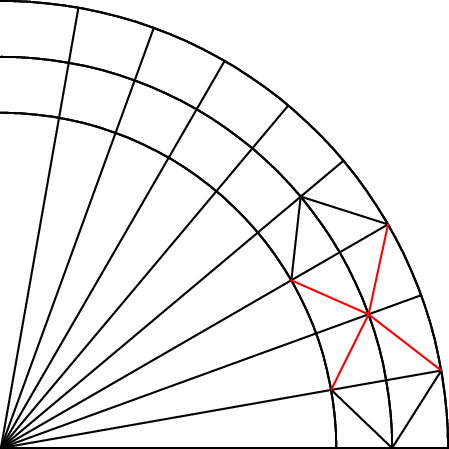}\\
\caption{Construction of ``Union Jack" mesh
by quadrilateral cell subdivision.}
\label{fig:uj1}
\end{figure}

Suppose that a conical surface is facetted with triangles
beyond a certain radius, in the following way, viz.\ it is
divided into four-sided areas by circles of increasing
radius centred on the cone apex
and by a set of straight lines directed radially outward from the apex.
These small areas are then bisected by lines joining their opposite
corners in an alternating pattern to produce a so-called
``Union Jack" mesh of triangles, see \Fig{uj1}.

Assume the cone to have apex angle~$\pi/2-\alpha$, so that its equation in 
cylindrical polar coordinates is $Z=-R \tan \alpha$.
Then if a triangle with its top corner at~$(R,0,Z)$
and node separation~$(\Delta r,0,\Delta z)$
has Cartesian coordinates of two of its nodes given by
\begin{equation}\label{eq:node2}
(x,y,z)=(R,0,Z),\;\;(R-\Delta r,0,Z-\Delta z)
\end{equation}
so that the third node is
\begin{equation}\label{eq:node3}
(x,y,z)=(R \cos \Delta \phi, R \sin \Delta \phi, Z)
\end{equation}
where $\Delta \phi$ is the angular separation between the radial mesh lines.
The vectors of the two sides meeting at the top of the triangle are thus
\begin{equation}\label{eq:trivec}
{\bf s}_1 = (\Delta r,0,\Delta z),\;\;{\bf s}_2=(R \cos \Delta \phi -R + \Delta r, R \sin \Delta \phi, \Delta z)
\end{equation}
and the unit normal to the triangle is given by
\begin{equation}\label{eq:trin}
\hat{\bf n}= \frac{{\bf s}_1 \times {\bf s}_2}{|{\bf s}_1 \times {\bf s}_2|}
\end{equation}
Straightforward calculation gives
\begin{equation}\label{eq:triun}
{\bf s}_1 \times {\bf s}_2= R \Delta r \left( \tan \alpha \sin \Delta \phi, -R(1-\cos \Delta \phi) \tan \alpha, \sin \Delta \phi \right)
\end{equation}
Substituting \Eq{triun} in \Eq{trin} gives,
assuming that $\Delta \phi \ll 1$, 
\begin{equation}\label{eq:nappr}
\hat{\bf n}\approx \left(\sin \alpha, \frac{\Delta \phi}{2}\sin \alpha, \cos \alpha \right)
\end{equation}

The barycentre of a small triangle lying in a curved surface 
may be assumed to coincide with that of its facetted approximation,
which lies at
\begin{equation}\label{eq:tric}
\frac{1}{3}\left( 2R -\Delta r+ R \cos \Delta \phi, R \sin \Delta \phi, 3Z -\Delta z \right)
\end{equation}
The formula for a unit normal at a point at angle~$\phi$
on the cone is
\begin{equation}\label{eq:nexac}
\hat{\bf n}= \left(\sin \alpha \cos \phi, \sin \alpha \sin \phi, \cos \alpha \right)
\end{equation}
Hence the exact normal at $\phi=\Delta \phi/3 \ll 1$ has components
\begin{equation}\label{eq:nexap}
\hat{\bf n}\approx \left(\sin \alpha, \frac{\Delta \phi}{3}\sin \alpha, \cos \alpha \right)
\end{equation}
verifying the formulae given in \Sec{surfn}.

There is the important subtlety that to re-use previous software, it
is convenient to work in the reordered coordinate system $(R,Z,\zeta)$,
where $\zeta$ is toroidal angle. However, to preserve system handedness,
it follows that $\zeta=-\phi$.

\section{Runge-Kutta-Fehlberg Schemes}\label{sec:rkf}
Two slightly different schemes are used, depending whether flux
coordinates are used to solve the fieldline equation. 
Both schemes are third order RKF, and
in each the results from computing with two Runge-Kutta schemes of
different orders gives an estimate for the error in the
fieldline position at step~$(n+1)$ for input to the Shampine-Watts
time step contol algorithm.

\subsection{Autonomous RKF}\label{sec:rkfa}
For the update in flux coordinates, a scheme which is third order
only for autonomous ODEs (ie.\ field independent of integration variable)
was developed.
The second order scheme of the pair for updating a scalar position~$x$
in a field~$B$ is
\begin{eqnarray}\label{eq:RKF2}
x_1 = x^n + \Delta B^n \\ \nonumber
\bar{x}_1 = x^n + \Delta B_1 \\ \nonumber
x_2 = \frac{1}{2} (x_1 + \bar{x}_1 )
\end{eqnarray}
where $\Delta$ is the timestep set by Shampine-Watts,
the value of field at step~$n$ is $B^n=B(x^n)$, $B_1= B(x_1)$ and
overbar denotes an auxiliary value of position.
The corresponding third order scheme reuses the value of~$B$ calculated above, 
so that
\begin{eqnarray}\label{eq:RKF3}
x_0 = x^n + \frac{1}{3} \Delta B^n\\ \nonumber
\bar{x}_0 = x^n + \frac{2}{3} \Delta B_1\\ \nonumber
\bar{\bar{x}}_1 = \frac{1}{2} (x_0 + \bar{x}_0 ) \\ \nonumber
\bar{x}_2 = x^n +  \Delta \bar{\bar{B}}_1\\ \nonumber
x_3 = \frac{1}{2} (x_2 + \bar{x}_2 )
\end{eqnarray}
where $\bar{\bar{B}}_1= B(\bar{\bar{x}}_1)$.
If $|x_2-x_3|< \epsilon_r ||x||$ then the step is accepted, $x^{n+1}=x_3$,
$n$ is incremented and the integration proceeds ($||x||$
estimates the size of~$x$).
This scheme has the potentially useful property for
fieldline intersection tests, that $x^{n+1}$ always lies within the safety range
$(x_{min},x_{max})$ given by
\begin{equation}\label{eq:srange}
\left( \min{(x_1, \bar{x}_1,\bar{x}_2)},\max{(x_1, \bar{x}_1,\bar{x}_2)} \right)
\end{equation}

\subsection{Non-autonomous RKF}\label{sec:rkfn}
The following RKF~$2(3)$ scheme is taken from~\cite{hairernorsettwanner}.
The same second order scheme~\Eq{RKF2} 
as before is combined with the third order accurate advance
\begin{eqnarray}\label{eq:RKF3n}
\bar{x}_2 = x^n +  \Delta B_2\\ \nonumber
x_3 = \frac{1}{3} ( x_2 + 2 \bar{x}_2 )
\end{eqnarray}
As usual for RK schemes, an update for a vector equation is produced
simply by substituting vectors for the scalars~$x$ and~$B$
in \Eqs{RKF2}{RKF3n}.

For this scheme, even in the scalar case,
there is no known way to bound with precision the region in which
the new value~$x^{n+1}=x_3$ must lie. 

\bibliographystyle{IEEEtran}
\bibliography{waynes,misc,new,warv,neuts,cadpapers,mc}
\end{document}